\documentclass[a4paper]{article}

\usepackage{INTERSPEECH2020}
\usepackage{times}
\usepackage{epsfig}
\usepackage{graphicx}
\usepackage{float}
\restylefloat{table}
\usepackage{amsmath}
\usepackage{amssymb}
\usepackage{hyperref}
\usepackage{mathtools}
\usepackage[table, dvipsnames]{xcolor}
\usepackage[linesnumbered,ruled]{algorithm2e}

\usepackage{array} 
\newcolumntype{C}[1]{>{\centering\let\newline\\\arraybackslash\hspace{0pt}}m{#1}} 

\usepackage{multirow, hhline, colortbl}

\title{Automatic Quality Assessment for Audio-Visual Verification Systems.\\ The \textit{LOVe} submission to NIST SRE Challenge 2019}
\name{Grigory Antipov, Nicolas Gengembre, Olivier Le Blouch, Ga\"el Le Lan}
\address{Orange Labs, 4 rue du Clos Courtel, Cesson S\'evign\'e, France}
\email{\{grigory.antipov,nicolas.gengembre,olivier.leblouch,gael.lelan\}@orange.com}

\begin{document}

\maketitle

\begin{abstract}

Fusion of scores is a cornerstone of multimodal biometric systems composed of independent unimodal parts. In this work, we focus on quality-dependent fusion for speaker-face verification. To this end, we propose a universal model which can be trained for automatic quality assessment of both face and speaker modalities. This model estimates the quality of representations produced by unimodal systems which are then used to enhance the score-level fusion of speaker and face verification modules. We demonstrate the improvements brought by this quality-dependent fusion on the recent NIST SRE19 Audio-Visual Challenge dataset.

\end{abstract}

\section{Introduction}

Accurate person recognition in videos is a holy grail of biometrics. Evaluation campaigns and datasets were organized to tackle this problem for applications such as indexation of TV content~\cite{galibert2013first} or secure bimodal authentication~\cite{messer1999xm2vtsdb}. That task had been rarely addressed for uncontrolled data, due to the lack of a challenging bimodal dataset, until the last Speaker Recognition Evaluation (SRE) organized by National Institute of Standards and Technology (NIST). The $2019$'s edition proposes an Audio-Visual challenge \cite{sadjadi20202019}, for which participants are allowed to use both speech and image signals to detect persons of interest in social media format videos.

Multimodal biometric fusion can be performed at various levels: representation, score or decision~\cite{ross2003information}. Due to the variety of representations and classifiers depending on each biometrics, it is usually easier to fuse at the score level. Score-level fusion strategies usually involve beforehand normalization making scores to have similar statistics~\cite{jain2005score}. Rapidly, the problem of degraded input data for some biometric modalities led to integrate quality information into the fusion strategy~\cite{fierrez2005discriminative}, where fingerprint images qualities were annotated by human experts.

Kimura et al.~\cite{kimura2014quality} employed quality features for score-level fusion of face, gait and height biometrics, using spatial and temporal resolution as quality estimates. Logistic regression was used for fusion, and the authors used a Gaussian process regression with a nonlinear kernel function to predict optimal fusion weights from qualities.
More generally, some previous works~\cite{ferrer2012unified,brummer2010abc} proposed quality estimates based on the hand-crafted \textit{expert features} extracted from the input signal. On the contrary, in this work, we propose a trainable model to automatically estimate a quality based on a face or speaker feature representation.
More precisely, our contributions are as follows: (1) we propose a universal model which is separately trained for quality estimation of speaker and face modalities, (2) we design $2$ independent unimodal systems for speaker and face verification, respectively, and (3) we apply our quality assessment model to improve the fusion of the unimodal verification systems on the NIST SRE19 Audio-Visual challenge dataset.

The rest of the paper is organized as following. Section~\ref{sec:related_works} is dedicated to related works. 
Section~\ref{sec:quality} details our quality assessment system and the corresponding quality-based fusion. Sections~\ref{sec:face_verification_pipeline} and \ref{sec:speaker_verification_pipeline} present face and speaker recognition pipelines, respectively. Section~\ref{sec:fusion_results} experimentally evaluates our systems, and Section~\ref{sec:conclusion} concludes the work.

\section{Related Works}
\label{sec:related_works}

\subsection{Face and Speaker Embeddings}

Contemporary face and speaker verification systems have similar structure.
Indeed, in both cases, Neural Networks (NNs) are trained to extract feature representations (commonly referred to as \textit{embeddings}) of the respective input modalities. For face verification, a large panoply of approaches has been proposed using Convolutional Neural Networks (CNNs)~\cite{taigman2014deepface,sun2014deep,wen2016discriminative,wang2018additive,deng2019arcface}, the current state-of-the-art (SOTA) being obtained by training with CosFace~\cite{wang2018additive} and ArcFace~\cite{deng2019arcface} losses.
Similarly, the SOTA embeddings for speaker recognition are also neural-based and known as \textit{x-vectors}~\cite{snyder2018x}.

However, the main difference between face and speaker verification is in the way these embeddings are compared to compute the similarity between the respective face images or speech recordings.
The two cases are detailed below.

At the dawn of the CNN-based face verification various methods have been explored to similarity computation, such as weighted $\chi^{2}$, Siamese networks~\cite{taigman2014deepface} or Joint Bayesian~\cite{sun2014deep}. As of today, the prevailing approach is a direct cosine in the face embeddings space. Such approach is indeed effective when the input face images allow extraction of discriminative embeddings, but the recent work of Shi and Jain~\cite{shi2019probabilistic} showed that it falls short when the face embeddings are extracted from the low-quality or occluded face images.

Contrary to face verification embeddings, the so-called \emph{x-vectors} are almost never used without a Probabilistic Linear Discriminant Analysis~\cite{ioffe2006probabilistic} (PLDA) scoring backend to deal with intra- and inter-speakers variabilities. End-to-end approaches enabling cosine similarities of embeddings remain quite marginal, mostly because of language adaptation, but the issue is more and more investigated~\cite{bhattacharya2019generative}.
In this work, we experiment with a pipeline inspired by Snyder~\cite{snyder2019speaker} and enriched by two complementary scoring backends: standard PLDA and direct cosine similarities.

\subsection{Automatic Quality Assessment}

While to the best of our knowledge no previous work proposed a dedicated system for automatic quality estimation of various modality representations, some studies are closely related to our main contribution.
More precisely, the already mentioned work of Shi and Jain~\cite{shi2019probabilistic} showed that a pair of noisy face embeddings can be very close to each other in the embeddings space while being extracted from faces of different persons which leads to false positive errors.
Therefore, Shi and Jain argued that for face verification on challenging datasets, it is essential to estimate the \textit{uncertainty} of the embedding and propose to do it via probabilistic modeling.
The authors reported that their model's convergence is unstable and highly dependent on the hyper-parameters, but their approach illustrates that quality estimation, which could be seen as a form of uncertainty, can be directly derived from the embeddings.
In the similar spirit, a very recent work of Silnova et al.~\cite{silnova2020unified} successfully employed probabilistic uncertainty estimation for speaker diarization.

Otherwise, the aggregation models~\cite{yang2017neural,xie2018multicolumn} estimate relative weights for a set of input embeddings, to compute a weighted average. However, these weights cannot be considered as absolute embedding qualities, as they are estimated w.r.t. a given set of embeddings. In this work, we propose a simple model, inspired by the aggregation model of Xie and Zisserman~\cite{xie2018multicolumn}, which estimates the quality of embeddings independently. We apply this model for face and speaker modalities.

\begin{figure*}[t!] \centering
\begin{tabular}{c|c}
    \begin{minipage}{7cm}
    \centering
        \includegraphics[width=\linewidth]{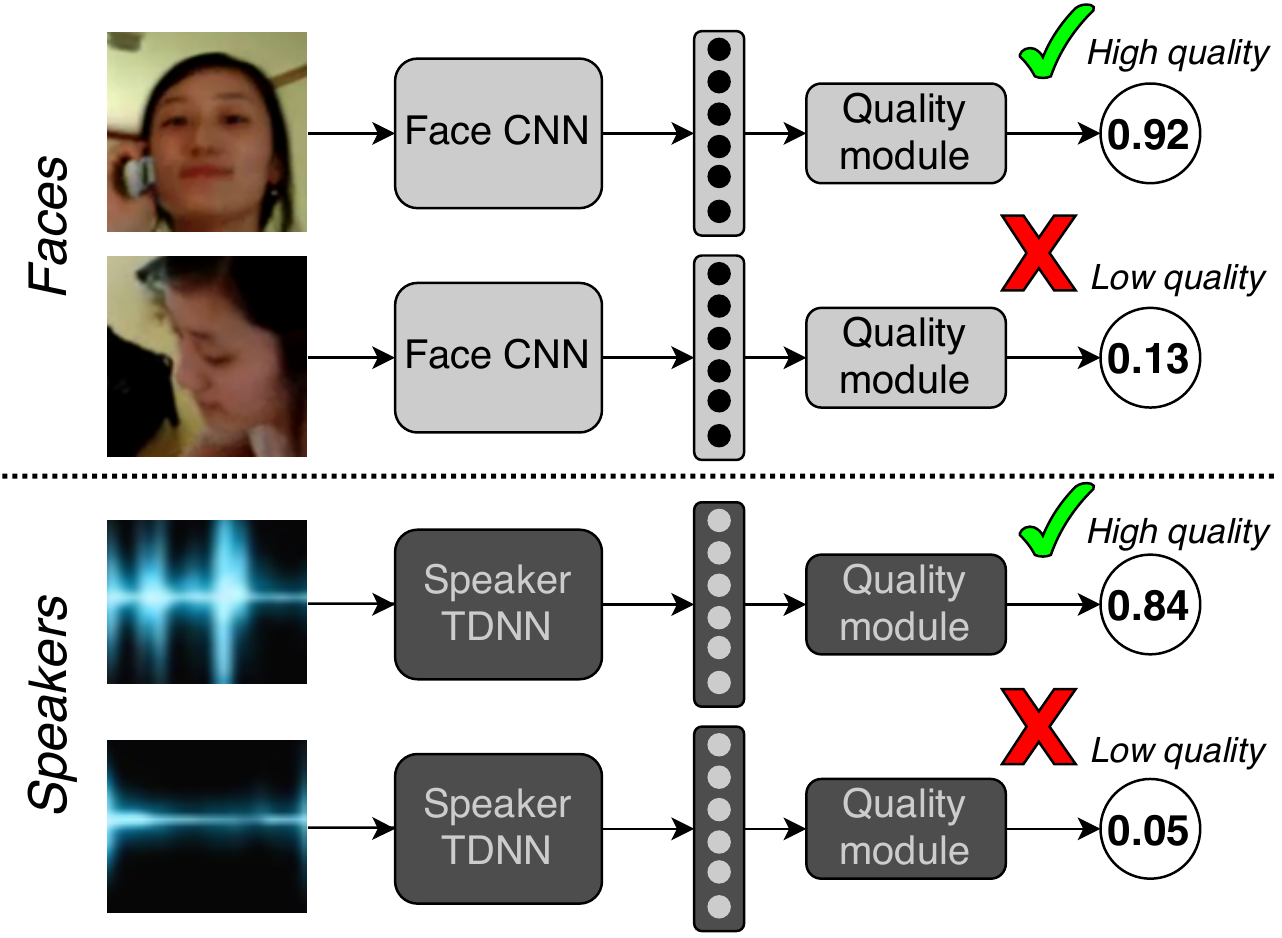} \\
        (a): inference
    \end{minipage}
    &
    \begin{minipage}{9cm}
        \centering
        \includegraphics[width=\linewidth]{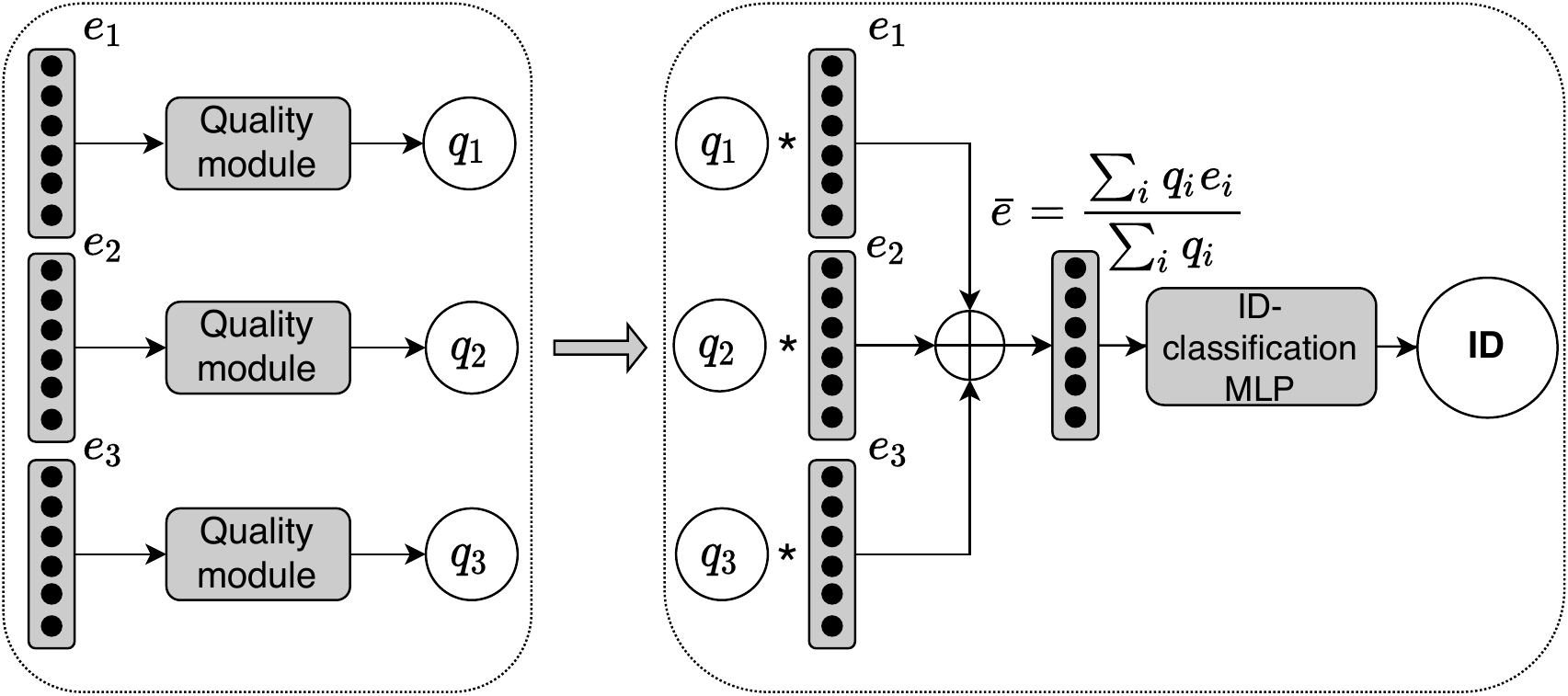} \\
        (b): training
    \end{minipage}
    \\
\end{tabular}
\caption{Proposed quality assessment modules for face and speaker embeddings. At inference time (a), the quality modules take the respective embeddings and output their real-valued qualities. A (face / speaker) quality module is trained (b) to optimize a weighted average embedding for further identity classification.}
\label{fig:qualityassessment}
\end{figure*}

\section{Quality and Score-Level Fusion}
\label{sec:quality}

\subsection{Quality Assessment}
\label{subsec:quality_assessment}

When fusing scores obtained from multiple modalities, one expects that the fusion algorithm learns the strengths and weaknesses of the unimodal algorithms.
For example, in an audio-visual verification task, such as the NIST SRE19 challenge, an ideal fusion approach would pay more attention to one of the two unimodal scores depending on the confidence of the audio and video systems (\textit{e.g.} it should pay more attention to the audio score, if the video signal is too dark and indistinguishable, and the other way around if the audio is recorded in a noisy environment).
Importantly, the unimodal scores themselves are often not sufficient to estimate the confidence of the respective predictions (indeed, as discussed in Section~\ref{sec:related_works}, a pair of blurry faces is often wrongly predicted as positive with a high verification score).
Therefore, in this Section, we propose a separate module for estimating the confidence of the unimodal systems.

Our quality assessment module is presented in Figure~\ref{fig:qualityassessment}.
This module is a neural network designed on the top of the (priorly trained) speaker / face embeddings extractor.
As depicted in Figure~\ref{fig:qualityassessment}-(a), at inference time, the embeddings (denoted  $e_i$) are the inputs of the quality assessment module which predicts a single real-valued $q_i$ (\textit{i.e.} the \textit{quality}) in the range $[0., 1.]$ by applying a sigmoid.

During training (\textit{cf.} Figure~\ref{fig:qualityassessment}-(b)), the quality module is optimized to find a weighted average of a set of $M$ embeddings ($M=3$ in the figure) belonging to the same person.
More precisely, the weights $q_{i}$ (\textit{i.e.} the qualities) of each embedding reflect to what extent the embedding participates in the linear combination:
\begin{equation}
\bar e = \frac{\sum\limits_{i=1\dots M}{q_i e_i}}{\sum\limits_{i=1\dots M}q_i}
\label{eq:qualembd}
\end{equation}
The resulting averaged embedding $\bar{e}$ is used for identity classification (the optimization is performed with the ArcFace loss~\cite{deng2019arcface}).
During this training process, the quality assessment module learns to assign larger weights to embeddings that are easier to classify, and lower weights to hardly classifiable ones.
Therefore, the weights learned this way implicitly represent the qualities of the embeddings.

\subsection{Quality-Based Score-Level Fusion}
\label{subsec:audiovisual_fusion}

The presented quality assessment module is used for the score-level fusion of audio and video scores in the following way.
For a single embedding, each of the 2 modalities generates 3 scalars, namely: the verification score $s$, the quality of the enrolment embedding $q_{e}$, and of the test one $q_{t}$.
Therefore, each \textit{trial} produces 6 scalars, namely: the 2 scores $s_{spk}$ and $s_{face}$ and the corresponding quality estimates $\mathbf{q_{spk}}=(q_{e},q_{t})_{_{spk}}$ and $\mathbf{q_{face}}=(q_{e},q_{t})_{_{face}}$.
These 6 values are fused via the quality-based score-level fusion with the Cllr-logistic regression~\cite{brummer2007fusion, kittler2007quality}, as shown in the following equation:

\begin{equation}
    LLR_{spk+face} = \sum_{i\in\{spk, face\}}a_i s_{i} + b_i q_{e_i} + c_i q_{t_i} + d
    \label{eq:lr_fusion}
\end{equation}
where $a_{i}$, $b_{i}$, $c_{i}$ and $d$ are the learned regression parameters.

Finally, it is important to notice, that the embedding qualities can be used not only for the inter-modality fusion as presented above, but also for fusion of scores inside a single modality.
Indeed, as it is common for challenges and as presented in Sections~\ref{sec:face_verification_pipeline} and~\ref{sec:speaker_verification_pipeline}, we propose several embedding extractors for each modality.
Therefore, the intra-modality verification scores can also be fused taking into account the qualities of the respective embeddings (\textit{cf.} Section~\ref{sec:fusion_results}).

\section{Face Verification Pipeline}
\label{sec:face_verification_pipeline}

In this Section, we present the visual part of our audio-visual system based on the faces detected in the video frames.
More precisely, video face verification is performed in 4 steps: 
Firstly, (1) the faces in video frames are detected and aligned.
Then (2) they are passed through an embedding extraction CNN.
After that, (3) the resulting face embeddings issued from various faces of a video should be combined (aggregated) taking into account the fact that they can potentially belong to different persons in the same video.
Finally, (4) the similarity between a trial of videos is estimated by computing the distance between the respective aggregated embeddings.

\subsection{Extraction of Face Embeddings}

We use SOTA standard approaches for steps (1) and (2).
Thus, face detection is performed at $1$ frame per second using the well-known MTCNN model~\cite{zhang2016joint} for detection and DLIB~\cite{dlib09} for alignment.
For face embeddings extraction, we have trained $4$ CNNs, namely ResNet-50~\cite{he2016deep}, PyramidNet~\cite{han2017deep}, ArcFace-50 and ArcFace-100~\cite{deng2019arcface} using CosFace and ArcFace losses.
All $4$ face embeddings extraction CNNs have been trained for the identity classification on a subset of the mixture of two public datasets, namely: MsCeleb1M~\cite{guo2016ms} and MegaFace~\cite{kemelmacher2016megaface}, containing about $7$M photos of $100$K classes (identities).
PyramidNet (the $101$-layers architecture has been chosen) and ResNet-50 have been trained with the CosFace loss, while ArcFace-50 and ArcFace-100 have been trained with the ArcFace loss.

\subsection{Aggregation of Face Embeddings}

Naturally, face embeddings extracted from different frames of a video are aggregated differently depending on whether the video is used for enrollment or test.
Indeed, enrollment videos have several annotated \textit{key} faces where the person to be enrolled is manually annotated with bounding boxes.
Therefore, the aggregation model serves to construct a single aggregated key embedding from the embeddings extracted from key faces.
For test videos, there is no prior on number of identities in a video.
Thus, we firstly cluster (agglomerative hierarchical clustering algorithm is employed) embeddings extracted in a video, and then aggregate them inside each obtained cluster.

We do not simply average face embeddings to aggregate them, but we rather train a separate $2$-layer Transformer-based model for this sake.
This model learns how to downplay the influence of poor quality embeddings and how to pay more attention to discriminative embeddings in order to minimize the classification loss.
The Transformer aggregation model is trained after the face embeddings extraction CNN (the latter being fixed during the training of the former) for the identity classification objective on the VGGFace2 dataset~\cite{cao2018vggface2}.
As a result, the usage of the Transformer-based aggregation instead of the trivial average aggregation improves the score of our face verification pipeline by more than $20$\% according to the challenge's metric.

\subsection{Calculation of Similarity between Face Embeddings}

In order to compute a similarity between a trial of enrollment and test videos $sim(enrol, test)$ (which is the ultimate step of our face verification pipeline), we calculate the similarities between the aggregated key enrollment embedding $\overline{x}_{enrol}^{(K)}$ and the aggregated embeddings of each test cluster $\overline{x}_{test}^{(c_{i})}$, and take the maximum: $sim(enrol,test)=\max\limits_{i}\{sim(\overline{x}_{enrol}^{(K)},\overline{x}_{test}^{(c_{i})})\}$.

The most common way of calculating a similarity between a pair of face embeddings is to simply use a cosine similarity: $sim(\cdot,\cdot)\equiv\cos{(\cdot,\cdot)}$.
Nevertheless, we have empirically observed that normalizing the cosine similarities with Adaptive-Symmetric NORMalization (\emph{as-norm})~\cite{cumani2011comparison} allows to largely avoid the false positive verification errors (which are the most costly ones according to the challenges' metrics).
Therefore, we employ \emph{as-norm} for score normalization in our final pipeline (the IJC-B dataset~\cite{maze2018iarpa} containing about $3500$ identities is used to calculate the reference cohort).

\begin{table*}[h!]
\centering
\begin{tabular}{|l|l|c|c|c|c|c|c|c|c|c|c|}
\hline
\multirow{2}{*}\textit{\#} & \multirow{2}{*}\textit{Fusion Rule} & \multicolumn{2}{c|}{\multirow{2}{*}\textit{Quality $q_{spk}$}} & \multicolumn{2}{c|}{\multirow{2}{*}\textit{Quality $q_{face}$}} &
\multicolumn{3}{c|}{\emph{SRE19-DEV}} & \multicolumn{3}{c|}{\emph{SRE19-EVAL}} \\
\hhline{~~----------}
&  & intra & inter & intra & inter & \textit{EER} & \textit{minC} & \textit{actC} & \textit{EER} & \textit{minC} & \textit{actC} \\
\hline
1 & Audio-only & & & & & 8.33 & 0.220 & 0.246 & 2.65 & 0.138 & 0.148 \\
2 & Visual-only & & & & & 8.33 & 0.179 & 0.186 & 2.21 & 0.066 & 0.069 \\
3 & Sum & & & & & 3.99 & 0.094 & 0.122 & 0.66 & 0.031 & 0.043 \\
\hline
4 & LR (\emph{official}) & & & & & 3.96 & \bf{0.087} & 0.101 & 0.66 & 0.039 & 0.042 \\
\hline
5 & LR & & & & & 3.99 & 0.094 & 0.101 & 0.66 & 0.032 & 0.034 \\
6 & LR & & \checkmark & & & 3.70 & 0.097 & 0.104 & \bf{0.60} & 0.032 & 0.035 \\
7 & LR & & & & \checkmark & \bf{2.78} & 0.094 & 0.097 & 0.88 & 0.027 & 0.027 \\
8 & LR & & \checkmark & & \checkmark & \bf{2.78} & 0.094 & 0.095 & 0.66 & 0.025 & 0.029 \\
\hline
9 & LR & \checkmark & \checkmark & \checkmark & \checkmark & \bf{2.78} & 0.088 & \bf{0.091} & 0.69 & \bf{0.021} & \bf{0.022} \\
\hline
\end{tabular}     
\caption{Audiovisual fusion results on SRE19-DEV and SRE19-EVAL. The columns $q_{spk}$ and $q_{face}$ indicate whether the respective qualities are used for intra- / inter-modality fusion (\textit{cf.} Section~\ref{subsec:audiovisual_fusion}). ``LR'' = Logistic Regression (Equation~\ref{eq:lr_fusion}). ``EER'' = Equal Error Rate. Row 4 reports our official Challenge primary result, based on an earlier version of the visual system, not presented in the paper.}
\label{tab:final_results_with_10_systems}
\end{table*}

\section{Speaker Verification Pipeline}
\label{sec:speaker_verification_pipeline}
This Section describes the global speaker verification pipeline. 
Firstly, (1) our two speaker identification systems are defined, followed by (2) the processing of test recordings by diarization and cluster selection, then (3) the experimental details.

\subsection{Speaker identification}

The two speaker identification systems in our pipeline are all based on a same \textit{x-vector} topology.
The first one respects the full Kaldi~\cite{povey2011kaldi} process described in~\cite{snyder2019speaker}. The second one is an end-to-end Pytorch~\cite{paszke2019pytorch} variant where the standard LLR is replaced by cosine similarity. This is achieved by replacing the top layers by an additional embedding layer, and refining the entire network with the ArcFace loss~\cite{deng2019arcface}.

\subsection{Diarization and cluster selection}

For test recordings, potentially containing several different speakers, an automatic \textit{diarization} step is mandatory to segment test recordings and extract speaker homogeneous clusters. That is where the procedure defined in~\cite{snyder2019speaker} is applied to our process. On test recordings, embeddings are extracted from 1.5 second segments with a 0.75 second overlap. 
LLRs or cosine similarities are then computed between all pairs of embeddings. 
After an agglomerative hierarchical clustering, we take the union of each of the individual diarization results in a set of ways to partition a recording that has at most $K$ speakers.
Finally, the computation of the trial score, or \textit{cluster selection} step, consists in taking the maximum of all scores obtained by an \emph{enrollment embedding} versus all the \emph{cluster embeddings} of a test recording (similarly to the face verification pipeline).

\subsection{Experimental details}
\label{subsec:speaker_pipeline_experiments}

The speaker identification systems are trained on the VoxCeleb dataset~\cite{nagrani2017voxceleb} with the Kaldi framework and Pytorch.
In order to increase the diversity of the acoustic conditions in the VoxCeleb dataset, a 8-fold augmentation strategy is used that adds seven corrupted copies of the original recordings to the training list. The recordings are corrupted by either digitally adding noise (i.e., babble, general noise, music), convolving with simulated room impulse response (RIR), high-pass and low-pass filtering, GSM-AMR and pitch/tempo shifting. 

For speech parameterization, we extract 30-dimensional MFCCs (including c0) from 25 ms frames every 10 ms using a 30-channel mel-scale filterbank spanning the frequency range 20 Hz – 7600 Hz. Before dropping the non-speech frames using a proprietary neural phonetic-based speech activity detection, a short-time cepstral mean subtraction is applied over a 3-second sliding window.
Our experiments led us to a full speaker verification pipeline resulting in a fusion of six versions of a similar process. 
They differ from the speaker identification systems used for the \emph{diarization} and \emph{cluster selection} steps by exploring complementary backend configurations.
For the sake of simplicity, these variants are not detailed in this paper.
The final submission for our audio system is a fusion of these six variants scores computed by Cllr-logistic regression~\cite{brummer2007fusion}, with a target prior of 0.05, which parameters were trained on SRE19-DEV.

\section{Fusion results}
\label{sec:fusion_results}

In this Section, we detail the score-level fusion results on both development and evaluation sets of challenge dataset (SRE19-DEV and SRE19-EVAL), for various fusion strategies, with or without 
quality estimates. Results are reported in Table \ref{tab:final_results_with_10_systems} in terms of $EER$, $min_C$ and $act_C$, as specified in \cite{sadjadi20202019}\footnote{According to the NIST SRE19 Challenge rules, we can only report our proper challenge results, but not the ones of other competitors.}.
The official 
SRE19 scoring toolkit was used to compute these metrics.

\subsection{Challenge Metrics}

A predefined detection cost function $C(\theta)$ (Equation \ref{eq:c_norm}), depending on the decision threshold $\theta$, allows to compute two metrics: $min_C$ and $act_C$ (the latter is the calibrated one).

\begin{equation}
\begin{cases}
C(\theta) = P_{FN}(\theta) + \frac{1-P_{tar}}{P_{tar}} \times P_{FP}(\theta)\\
min_C = \min_{\theta}(C)\\
act_C = C\left(\theta = \log\left(\frac{1-P_{tar}}{P_{tar}}\right)\right)
\end{cases}
\label{eq:c_norm}
\end{equation}

$P_{FP}(\theta)$ (resp. $P_{FN}(\theta)$) is the false positive (resp. false negative) probability given the threshold $\theta$. The target prior $P_{tar}$ is set by the organizers at 0.05. The $act_C$ corresponds to the cost when $\theta$ is Bayes-optimal~\cite{de1970optimal}. It is the reference metric of the challenge, which simulates performances of the system in operational conditions. This metric is much more sensitive to false positive errors than to false negative ones.

The NIST SRE19 Audio-Visual task is threshold-dependent. Therefore calibration is required. Before fusion, all scores are calibrated on SRE19-DEV (for which labels are available), using the \emph{pybosaris} implementation\footnote{https://gitlab.eurecom.fr/nautsch/pybosaris} of Cllr-logistic regression~\cite{brummer2007fusion} with a target prior of 0.05. As mentioned in Section \ref{subsec:audiovisual_fusion}, Cllr-logistic regression is also used for fusion (LR in table \ref{tab:final_results_with_10_systems}), with the same target prior.

\subsection{Fusion: Score Only - $baseline$}

The first 3 rows of Table~\ref{tab:final_results_with_10_systems} present the final speaker (row 1) and face (row 2) recognition performances and those obtained when applying the sum rule on speaker and face LLRs~\cite{kittler1998combining} (row 3). These first results show that the audio and visual systems are relatively well calibrated with a gap between $min_C$ and $act_C$ of 0.026 in the worst case (speaker system on SRE19-DEV). It can also be noted that the task of person verification is harder on SRE19-DEV than on SRE19-EVAL, since better performances are obtained on SRE19-EVAL overall.
The visual system gives better performances than the audio one with a $min_C$ of 0.179 vs. 0.220 on SRE19-DEV (resp. 0.066 vs. 0.138 on SRE19-EVAL). Applying a simple sum rule shows both modalities are complimentary since the $min_C$ drops to 0.094 (resp. 0.031), while the $EER$ drops from 8.33\% to 3.99\% (resp. from (2.65\%, 2.21\%) for both modalities to 0.66\%). Replacing the sum rule by the Cllr-logistic regression (LR in the Table, row 5) does not change $EER$ and $min_C$ results but improves $act_C$ performances, as this formula calibrates the fused LLRs. 

\subsection{Quality-based fusion}

Rows 6-9 of Table \ref{tab:final_results_with_10_systems} present quality-based fusion performances. Qualities are included in the Cllr-logistic regression, as explained in Section \ref{subsec:audiovisual_fusion}.

Results show that incorporating quality estimates into the logistic regression improves all evaluation metrics. Contrastive experiments are proposed (rows 6 \& 7), where the quality of only one modality is provided. This shows that the proposed quality estimation works better for face embeddings than for speaker embeddings. However, incorporating both quality estimates is complementary (row 8). Finally, row 9 of Table \ref{tab:final_results_with_10_systems} presents an experiment where quality estimates are also used for the fusion of within-modality system variants : this approach gives the best results overall.

\section{Conclusion}
\label{sec:conclusion}

We proposed a universal automatic quality estimation module integrated in a multimodal system for audio-visual person verification.
We firstly presented our face and speaker verification pipelines. Then we demonstrated that our quality estimation module improves the fusion of the audio-visual scores both on the intra- and inter-modality levels. This was validated on the NIST SRE19 Audio-Visual Challenge.

\bibliographystyle{IEEEtran}
\bibliography{ms}

\end{document}